\documentclass[preprint2]{proto}
\usepackage{times}

\newcommand{\hide}[1]{}

\newcommand{\refs}{\par\noindent\hangindent=1pc\hangafter=1}
\begin{document}

\title{\textbf{\LARGE Collisional Formation and Modeling of Asteroid Families}}

\author {\textbf{\large Patrick Michel}}
\affil{\small\em Lagrange Laboratory, University of Nice-Sophia, CNRS, C\^ote d'Azur Observatory}

\author {\textbf{\large Derek C. Richardson}}
\affil{\small\em Department of Astronomy, University of Maryland}

\author {\textbf{\large Daniel D. Durda}}
\affil{\small\em Southwest Research Institute}

\author {\textbf{\large Martin Jutzi}}
\affil{\small\em University of Bern, Center for Space and Habitability, Physics Institute}

\author {\textbf{\large Erik Asphaug}}
\affil{\small\em School of Earth and Space Exploration, Arizona State University}

\begin{abstract}
\begin{list}{ } {\rightmargin 1in}
\baselineskip = 11pt
\parindent=1pc
{\small In the last decade, thanks to the development of sophisticated numerical codes, major breakthroughs have been achieved in our understanding of the formation of asteroid families by catastrophic disruption of large parent bodies. In this review, we describe numerical simulations of asteroid collisions that reproduced the main properties of families, accounting for both the fragmentation of an asteroid at the time of impact and the subsequent gravitational interactions of the generated fragments. The simulations demonstrate that the catastrophic disruption of bodies larger than a few hundred meters in diameter leads to the formation of large aggregates due to gravitational reaccumulation of smaller fragments, which helps explain the presence of large members within asteroid families. Thus, for the first time, numerical simulations successfully reproduced the sizes and ejection velocities of members of representative families. Moreover, the simulations provide constraints on the family dynamical histories and on the possible internal structure of family members and their parent bodies. \\~\\~\\~}

\end{list}
\end{abstract}

\section{\textbf{INTRODUCTION}}

Observed asteroid families in the main asteroid belt are each composed of bodies that are thought to originate from the catastrophic disruption of larger parent bodies (e.g., {\em Farinella et al.}, 1996). Cratering collisions can also lead to families, such as the one associated with asteroid Vesta, but we do not address this origin scenario here as few families have been linked to cratering events and their modeling requires a different approach (see the chapter by Jutzi et al. in this volume for more details). A few tens of asteroid families have been identified, corresponding to groups of small bodies well-concentrated in proper-orbital-element space (see, e.g., {\em Hirayama} 1918, {\em Arnold} 1969, and the chapter by Nesvorn\'y et al.\ in this volume) and sharing similar spectral properties (see, e.g., Chapman et al.\ 1989 and the chapter by Masiero et al.\ in this volume). Large families contain up to several hundred identified members, while small and compact families have of the order of ten identified members. Interestingly, the theory of the collisional origin of asteroid families rested for decades entirely on these similarities in dynamical and spectral properties and not on the detailed understanding of the collisional physics itself. Indeed, laboratory experiments on centimeter-scale targets, analytical scaling rules, or even complete numerical simulations of asteroid collisions were unable to reproduce the physical and dynamical properties of asteroid families (e.g., {\em Ryan and Melosh}, 1998). The extrapolation of laboratory experiments to asteroidal scales yielded bodies much too weak to account for both the size distribution and the dynamical properties of family members. In other words, there was no solution to match both the sizes and ejection velocities of family members simultaneously. To produce the large (assumed coherent) fragments seen in real families required an impact energy leading to ejection speeds of individual fragments that were much too small for them to overcome their own gravitational attraction. The parent body would then be merely shattered but not dispersed and therefore no family would be created ({\em Davis et al. 1979}). Conversely, matching individual ejection velocities and deriving the necessary fragment distribution resulted in a size distribution in which no big fragment was present, contrary to most real families (e.g., {\em Davis et al.}, 1985, {\em Chapman et al.}, 1989). Thus, big sizes implied no fragment dispersion (at the level of the dispersion of family members), and fragment dispersion implied no big fragments (at the level of the sizes of family members). 

A big caveat in the extrapolation of laboratory results to large asteroid scales is that the role of gravity (of both the targets and their fragments) is not taken into account in a laboratory-scale disruption involving cm-size targets. Thus, the role of gravity in the catastrophic disruption of a large asteroid at the origin of a family remained to be assessed. Indeed, a possible scenario reconciling the sizes and ejection velocities of family members could be that the parent body (up to several hundred kilometers in size) is disrupted into small pieces by the propagation of cracks resulting from a hypervelocity impact but then the small fragments generated this way would typically escape from the parent and, due to their mutual gravitational attraction, reaccumulate elsewhere in groups in order to build up the most massive family members. This idea had been suggested previously by {\em Chapman et al.} (1982), and numerical simulations by {\em Benz and Asphaug} (1999) had already shown that at least the largest remnant of an asteroid disruption had to be a bound aggregate. However, the formation of a full family by reaccumulation of smaller fragments remained to be demonstrated.

In the last decade, the formation of asteroid families was simulated explicitly for the first time accounting for the two phases of a large-scale disruption: the fragmentation phase and the gravitational reaccumulation phase. In these simulations, the two phases are usually computed separately using a hybrid approach. This chapter reviews the major advances achieved since {\em Asteroids III} thanks to this new modeling work, and the implications for our understanding of asteroid family formation and properties. 

\section{\textbf{SIMULATING A FAMILY-FORMING EVENT}}

Families are thought to form from the disruption of a large asteroid, called the parent body, as a result of the impact of a smaller projectile. Simulating such a process requires accounting for both the propagation of cracks in the parent body, leading to its conversion into separate fragments, and the possible gravitational interactions of these fragments. As explained above, the latter gravitational phase turns out to be crucial for reproducing asteroid family properties. In asteroid disruptions resulting from an hypervelocity impact, the fragmentation and the gravitational reaccumulation phases have very different associated dynamical times. In the fragmentation phase, the time scale for the propagation of the shock wave is determined by the target's diameter divided by the sound speed of the material (a few to tens of seconds for an asteroid 100 km in diameter). In the second phase, the time scale for gravitational reaccumulation is proportional to $1/\sqrt{G\rho}$, where $G$ is the gravitational constant and $\rho$ is the target bulk density, which corresponds to at least hours for $\rho$ = 1 to 3 g cm$^{-3}$. Therefore it is possible to model the collisional event by separating the two phases. A hybrid approach is generally adopted that consists of simulating first the fragmentation phase using an appropriate fragmentation code (called a hydrocode; see the chapter by {\em Jutzi et al.} in this volume), and then the gravitational phase, during which the fragments produced by the fragmentation phase can interact under their mutual attraction, using a gravitational $N$-body code. 

\bigskip
\noindent
\textbf{2.1 The Fragmentation Phase}
\bigskip

Several hydrocodes exist and are used in the planetary science community (see the chapters by {\em Asphaug et al.} and {\em Jutzi et al.} in this volume). In the first studies devoted to direct simulations of asteroid family formation ({\em Michel et al.}, 2001, 2002, 2003, 2004), a three-dimensional ``smoothed-particle hydrodynamics'' (SPH) code was used. This code solves in a Lagrangian framework the usual conservation equations (mass, momentum, and energy) in which the stress tensor has a nondiagonal part. The first families modeled in this way (Eunomia, Koronis and Flora) were of S taxonomic type. S-type asteroids are expected to be mostly made of ordinary chondrite materials, for which basalt plausibly has similar properties and therefore the parent bodies of these S-type families were assumed to be non-porous basalt. The Tillotson equation of state for basalt was used ({\em Tillotson}, 1962), which is computationally expedient while sophisticated enough to allow its application over a wide range of physical conditions. Plasticity was introduced by modifying the stresses beyond the elastic limit with a von Mises yielding relation ({\em Benz and Asphaug}, 1994, 1995). A yielding relation accounting for the dependence of shear stress on pressure, such as the Mohr-Coulomb or Drucker Prager ones, is generally more appropriate for rock material  (see the chapter by {\em Jutzi et al.} in this volume). It turns out to be important for cratering events for which part of the process is dominated by shearing or when the impacted body is composed of interacting boulders (a so-called ``rubble pile'' ({\em Davis et al.}, 1979) or gravitational aggregate ({\em Richardson et al.}, 2002); see also {\em Jutzi}, 2015). However, it was found that in the case of the disruption of monolithic parent bodies, the details of the strength model (e.g., pressure-dependent vs. pressure-independent yield strength) do not play an important role ({\em Jutzi} 2015). For the lower tensile stresses associated with brittle failure, a fracture model was used, based on the nucleation of incipient flaws whose number density is given by a Weibull distribution ({\em Weibull}, 1939, {\em Jaeger and Cook}, 1969). {\em Durda et al.} (2004) and {\em Nesvorn\'y et al.} (2006) used a similar hydrocode to study the formation of satellites from asteroid disruptions and other family-forming events. {\em Leinhardt and Stewart} (2009) studied large-scale disruptions and modeled the shock deformation with an Eulerian shock-physics code, CTH ({\em McGlaun et al.}, 1990), instead of the Lagrangian SPH code used in previous works. 

Recently, the SPH impact code used by {\em Michel et al.} (2001, 2002, 2003, 2004) was extended to include a model adapted for microporous materials ({\em Jutzi et al.}, 2008, 2009, this volume). The formation of asteroid families formed from a microporous parent body, such as for dark- (carbonaceous) type families, could thus also be investigated ({\em Jutzi et al.}, 2010, {\em Michel et al.,} 2011). Another study looked at the case of rubble-pile parent bodies (containing macroporous voids; {\em Benavidez et al.}, 2012). 

\bigskip
\noindent
\textbf{2.2 The Gravitational Phase}
\bigskip

Once the fragmentation phase is over and fractures cease to propagate (within the first simulated tens of seconds), the hydrodynamic simulations are stopped and intact fragments are identified. For impact energies typical of asteroid disruptions and for targets with a diameter typically greater than 1 km, it was found that the bodies are totally shattered into fragments of mass equal to the mass resolution of the simulations. In the first simulations performed in this way, the numerical resolution was limited to a few $10^5$ particles and corresponded to minimum boulder diameters of about 1 to 4 km, for a parent body of a few hundred kilometers in diameter. Thanks to increased computer performance, it is now possible to perform simulations with up to several million particles. However, the gain in particle size resolution is not dramatic and simulations are still limited to minimum fragment diameters of a few hundred meters for target diameter of a few hundred kilometers. Reaching fragment diameters down to meters or less is beyond the capabilities of current and probably near-future technologies. Only when the target's diameter is in the few hundred meters range can this minimum size be reached, but unfortunately, no asteroid family can be identified involving a parent body of such a small size. 

Once identified in the simulation outcome, the fragments and their corresponding velocity distributions are then fed into a gravitational $N$-body code, which computes the gravitational evolution of the system from the handoff point to subsequent hours or days of simulated time. Because the number of fragments is up to a few $10^6$, and their gravitational interaction as well as their potential collisions need to be computed over long periods of time (up to several simulated days), a very efficient $N$-body code is required to compute the dynamics. The most appropriate code to tackle this problem, which is the only one used so far by various groups to simulate the outcome of the gravitational phase of a collision, is the code called {\em pkdgrav} (see {\em Richardson et al.}, 2000 for the first application of this code to solar system problems). This parallel hierarchical tree code was developed originally for cosmological studies.  Essentially, the tree component of the code provides a convenient means of consolidating forces exerted by distant particles, reducing the computational cost, with the tradeoff of introducing a slight force error (of order 1\%) that does not affect the results appreciably since the dynamics are dominated by dissipative collisions. The parallel component divides the work evenly among available processors, adjusting the load at each timestep according to the amount of work done in the previous force calculation. The code uses a straightforward second-order leapfrog scheme for the integration and computes gravity moments from tree cells to hexadecapole order.

For the purpose of computing the gravitational phase of an asteroid disruption during which the generated fragments can interact and collide with each other, collisions are identified at each step with a fast neighbor-search algorithm in {\em pkdgrav}. Once a collision occurs, because the relative speeds are small enough (of the order of meters per second), it is assumed that no further fragmentation takes place between components generated during the fragmentation phase. In fact, the simulations presented by {\em Michel et al.} (2001) assumed perfect sticking of colliding fragments and all colliding fragments were forced to merge into a single particle regardless of their relative velocities. This assumption is justified because the initial impact results in an overall expanding cloud of fragments of relatively small individual masses, down to the minimum fragment size imposed by the numerical resolution, and colliding fragments have typical relative speeds that are smaller than their individual escape speeds. Since the fragments in {\em pkdgrav} are represented by spheres, when two spherical fragments reacumulate, they are merged into a single spherical particle with the same momentum. The same assumption was used by {\em Durda et al.} (2004) and {\em Nesvorn\'y et al.} (2006) in their studies of satellite formation and other family forming events. In a second and subsequent papers, {\em Michel et al.} (2002, 2003, 2004) improved their treatment of fragment collisions by using a merging criterion based on relative speed and angular momentum. In this case, fragments are allowed to merge only if their relative speed is smaller than their mutual escape speed and the resulting spin of the merged fragment is smaller than the threshold value for rotational fission (based on a simple prescription of a test particle remaining on the equator of a sphere). Non-merging collisions are modeled as bounces between hard spheres whose post-collision velocities are determined by the amount of dissipation taking place during the collisions. The latter is computed in these simulations using coefficients of restitution in the normal and tangential directions (see {\em Richardson}, 1994, for details on this computation). Note that {\em Durda et al.} (2011, 2013) performed bouncing experiments between 1-meter granite spheres as well as between cm-scale rocky spheres. These experiments gave a value for the normal coefficient of restitution of $\approx 0.8$, although much lower values are found with increasing roughness of contact surfaces. These results are particularly interesting because they are performed in an appropriate size regime (meter-sized bodies). However, bouncing in simulations occur at somewhat higher speeds (up to tens of m/s) than in those experiments, which may result in a decrease in the coefficient of restitution due to the start of cracking and other energy dissipation processes. Moreover, although in our numerical modeling, perfect spheres are used, it is reasonable to account for actual irregularities of fragments formed during the fragmentation phase to set the value of the coefficient of restitution. Since the values of these coefficients are poorly constrained, we usually arbitrarily set them equal to 0.5, meaning, for example, the rebound speed is set to half the impact speed.

More recently, {\em Richardson et al.} (2009) enhanced the collision handling in {\em pkdgrav} to preserve shape and spin information of reaccumulated bodies in high-resolution simulations of asteroid family formation. Instead of merging, fragments are able to stick on contact and optionally bounce or subsequently detach, depending on user-selectable parameters that include for the first time several prescriptions for variable material strength/cohesion. As a result, the reaccumulated fragments can take a wide range of shapes and spin states, which can be compared with those observed. This comes with a cost in terms of computation time as several weeks to months are needed for one simulation using a few tens of current processors. This is the reason why this approach has so far only been used for particular cases, such as modeling the formation of the asteroid Itokawa ({\em Michel and Richardson}, 2013), and not systematically for family formation investigations.

And finally, we must note that the Soft-Sphere Discrete Element Method (SSDEM) has been introduced in {\em pkdgrav} ({\em Schwartz et al.}, 2012, chapter by {\em Murdoch et al.} in this volume), which accounts more realistically for the contact forces between colliding/reaccumulating particles. This method should eventually replace the one developed by {\em Richardson et al.} (2009) to investigate the shape of reaccumulated fragments as it avoids arbitrary particle sticking and rather let the reaccumulated particles evolve naturally towards the resulting equilibrium shape of the aggregate. However, solving for all contact forces between particles over the whole time scale of the gravitational phase, and covering a large enough parameter space (accounting for the uncertainty on the various friction coefficients) remains a computational challenge. Nevertheless, some collisional studies started using the SSDEM implementation in {\em pkdgrav} focusing on low-speed impact events. In effect, no fragmentation code was used for the impact phase, which is needed for impacts during which the sound speed of the material is reached. Thus, {\em Ballouz et al.} (2014a, b) used {\em pkdgrav} and SSDEM to simulate low speed impacts between rotating aggregates and to investigate the influence of the initial rotation of colliding bodies on the impact outcome and the sensitivity of some friction parameters. The number of particles was small enough ($10^4$ at most) that simulations could be performed within a reasonable computation time. In the case of a family formation, the outcome of hydrocode simulations consists in several hundred thousands to millions of particles. Feeding them into the SSDEM version of {\em pkdrgav} requires another level of computer performance, although tests are under way.

\section{\textbf{MODELING THE FAMILY PARENT BODIES}}

Different possible internal structures have been considered for the family parent bodies. Monolithic parent bodies composed of one material type with or without microporosity (meaning micropores in the solid rock; see the chapter by {\em Jutzi et al.} in this volume, for a definition of microporosity) have been considered, as well as pre-shattered or rubble-pile parent bodies, with or without microporosity in the solid components. The assumed pre-shattered state could be seen as a natural consequence of the collisional evolution of main-belt asteroids. Indeed, several studies (see, e.g., {\em Asphaug et al.}, 2002; {\em Davis et al.}, 2002; {\em Richardson et al.}, 2002) have indicated that for any asteroid, collisions at high impact energies leading to a disruption occur with a smaller frequency than collisions at lower impact energies leading to shattering effects only. Thus, in general, a typical asteroid gets battered over time until a major collision eventually disrupts it into smaller dispersed pieces. Consequently, since the formation of an asteroid family corresponds to the ultimate disruptive event of a large object, the internal structure of this body before its disruption may be shattered by all the smaller collisional events that it has suffered over its lifetime in the belt, as suggested by {\em Housen} (2009) based on laboratory experiments and extrapolations using scaling laws. This would result in the presence of internal macroscopic damaged zones and/or voids. 

To model a pre-shattered target, {\em Michel et al.} (2003, 2004) devised an algorithm that distributes a given number of internal fragments of arbitrary shape and size within the volume of the parent body. The reason the internal fragments are given arbitrary shapes is that a network of fractures inside a parent body resulting from many uncorrelated small impacts is unlikely to yield spherical internal fragments whose sizes follow a well-defined power law. Then, void spaces are created by randomly removing a given number of particles from the fractured set. Since there are various ways to define a pre-shattered internal structure, {\em Michel et al.} (2004) also built a model of a pre-shattered parent body in which large fragments are preferentially distributed near the center and smaller fragments are generated close to the surface. 

Another model, closer to the definition of a rubble pile was also built by {\em Michel et al.} (2003). In that case, spherical components whose sizes followed a specified power law distribution were distributed at random inside the parent body. Particles not belonging to one of these spherical components were removed to create void space and particles at the interface of two or more spherical components were assigned to fractures. Some simulations were performed using those two additional models and the collisional outcomes did not show any major qualitative difference compared to those obtained from the first pre-shattered model.  {\em Benavidez et al.} (2012) constructed arbitrary rubble-pile targets by filling the interior of a 100-km-diameter spherical shell with an uneven distribution of solid basalt spheres having diameters between 8 km and 20 km. However, simulations performed so far using such rubble-pile parent bodies used a version of a SPH hydrocode with a strength model that did not allow the proper modeling of friction between the individual components of the rubble pile. As found by {\em Jutzi} (2015), the bodies in this case show a fluid-like behavior and are very (somewhat unrealistically) weak. Therefore in the following, we will only consider the results obtained for monolithic and pre-shattered bodies (as defined by {\em Michel et al.,} 2003, 2004) either with or without microporosity.

\section{\textbf{REPRODUCING WELL-KNOWN FAMILIES}}

For the first time, {\em Michel et al.} (2001) simulated entirely and successfully the formation of asteroid families from monolithic basalt-like parent bodies in two extreme regimes of impact energy leading to either a small or a large mass ratio of the largest remnant to the parent body $M_{lr}/M_{pb}$. Two well-identified families were used for comparison with simulations: the Eunomia family, with a 284~km diameter parent body and $M_{lr}/M_{pb} \approx 0.67$, was used to represent the barely disruptive regime, whereas the Koronis family, with a 119~km diameter parent body and $M_{lr}/M_{pb} \approx 0.04$, represented the highly catastrophic one. Both families are of S taxonomic type, for which ordinary chondrites are the best meteorite analog, but basalt material is typically used as analog material in collisional studies. In these simulations, the collisional process was carried out to late times (typically several days), during which the gravitational interactions between the fragments could eventually lead to the formation of self-gravitating aggregates (as a representation of rubble piles) far from the largest remnant. These first simulations assumed perfect sticking of reaccumulated fragments, regardless of relative speed and mass. This treatment was improved by {\em Michel et al.} (2002), allowing for the dissipation of kinetic energy in such collisions and applying an energy-based merging criterion, as described previously. This improved treatment did not change the conclusion obtained with the more simplistic method because typical relative speeds between ejected fragments are most often below their mutual escape speed. Therefore, this new set of simulations confirmed the idea that the reaccumulation process is at the origin of large family members. {\em Durda et al.} (2007) and {\em Benavidez et al.} (2012) made a systematic study of collisional disruption of monolithic and rubble-pile basalt-like 100 km-diameter parent bodies, assuming perfect sticking during reaccumulation, over a large range of impact conditions, and then re-scaled their results to compare with real families, showing again that the reaccumulation process is necessary to find any good solution. However, a caution about extending results from the disruption of 100 km-diameter parent bodies to observed families that originated from parent bodies very different in size from 100 km is in order and was acknowledged by {\em Durda et al.} (2007).

\bigskip
\noindent
\textbf{4.1 The Size Distribution of Family Members}
\bigskip

The role of geometric constraints accounting for the finite volume of the parent body in the production of family members was investigated by {\em Tanga et al.} (1999) and {\em Campo Bagatin and Petit} (2001). By filling the parent body with spherical ({\em Tanga et al.} 1999) or irregular ({\em Campo Bagatin and Petit} 2001) pieces, starting from the largest member, they were able to reproduce the size distribution of some asteroid families to an encouraging level of agreement. However, these models do not incorporate any physics, nor do they take into account fragment reaccumulation, therefore they do not provide any explanation for how family members are formed nor any prediction for their internal properties. Moreover, ejection velocities are not addressed by this approach.
 
The first full numerical simulations of catastrophic disruption and gravitational reaccumulation by {\em Michel et al.} (2001) assumed a monolithic structure of the parent body represented by a sphere with material properties of basalt (no internal porosity was considered). These simulations already reproduced qualitatively the main properties of real family member size distributions. However, when looking into more quantitative details, it was found that the cumulative size distribution of simulated fragments was characterized systematically by a lack of intermediate-sized bodies and a very steep slope for the smaller ones (see an example in Fig.~1). Such characteristics are not always observed in the size distributions of real family members. In fact, for some families, the size distribution looks rather continuous. In their systematic study, {\em Durda et al.} (2007), using the same internal structures (monolithic, basalt) but considering 100 km-diameter parent bodies only, found a larger variety of size distributions in terms of power-law slopes and discontinuities, depending on the considered impact conditions.

\begin{figure*}
 \epsscale{1.}
 \plotone{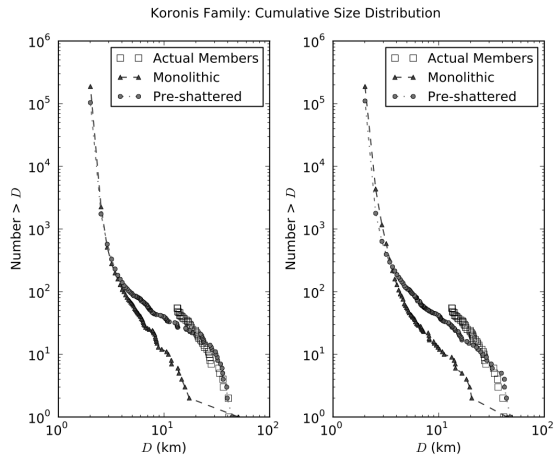}
 \caption{\small Cumulative diameter distributions in log-log plots for the fragments of the simulated Koronis families. The plot on the left was obtained with a
projectile colliding head-on, whereas an impact with an angle of incidence $\theta$ equal to $45^\circ$ gave rise to that on the right. Different symbols are used to
distinguish between parent-body models. The plots also show the estimated sizes of the actual members down to the completeness limit ({\em Tanga et al.}, 1999).
Note that the simulations using a pre-shattered target reproduce the four nearly identical largest members.}  
 \end{figure*}

Nevertheless, it is also known that the outcome of a collision is influenced by the initial internal structure of the parent body and that, depending on the initial structure, the fragment size distribution may be more or less continuous. In order to check this, {\em Michel et al.} (2003, 2004) modeled parent bodies with an internal structure composed of different zones of voids and fractures, as if they had first been shattered during their collisional history before undergoing a major event leading to their disruption, as explained in Sec.~3. The simulations of the Eunomia and Koronis family formations were redone using pre-shattered parent bodies ({\em Michel et al.}, 2004), and  the results were compared with those obtained using monolithic parent bodies. The best agreement was actually found with pre-shattered parent bodies. In particular, in the case of the Koronis family, an interesting result was obtained from these simulations, which may have important implications concerning the real family history. The size distribution obtained from the disruption of a pre-shattered parent body contains four largest fragments of approximately the same size, as can be seen in Fig.~1. This peculiar characteristic is shared by the real family, and has been a source of debate as it was assumed that a single collisional event cannot produce such a property (see {\em Michel et al.}, 2004, for a discussion). Moreover, the simulation using a monolithic parent body did not result in such a distribution. It was then demonstrated numerically for the first time, by using a pre-shattered parent body, that these fragments can actually be produced by the original event, and therefore no subsequent mechanism needs to be invoked to form them, which would otherwise require a revision of the entire family history ({\em Marzari et al.}, 1995). According to these results showing that even old families may well have originated from pre-shattered parent bodies, it was concluded that most large objects in the present-day asteroid belt may well be pre-shattered or self-gravitating aggregates/rubble piles.

Constraints provided by the measured size distribution of family members can eliminate formation scenarios in numerical simulations.
An interesting example is the Karin cluster, a small asteroid family identified by {\em Nesvorn\'y et al.} (2002) that formed $\sim$ 5.8 Myr ago in the outer main belt. The estimated size distribution for this family, when first identified, was fairly smooth and continuous over all sizes.  In particular, there was no big gap between the size of the largest member of the family, (832) Karin, and that of the second-largest member, (4507) 1990 FV. Numerical simulations by {\em Michel et al.} (2003) indicated that the best match to the continuous size distribution was provided by the break-up of a pre-shattered or rubble-pile parent body. Simulations starting from a monolithic parent body, on the other hand, produced size distributions showing a large gap between the sizes of the largest and next-largest fragments. The finding that the parent body of the Karin cluster needed to be a rubble pile was actually consistent with its history. Specifically, the parent asteroid of the Karin cluster is thought to have been produced by an early disruptive collision that created the much larger Koronis family some 2--3 Gyr ago. According to the results of Koronis family-formation simulations, the parent asteroid of the Karin cluster should have been formed as a rubble pile from Koronis family debris.

However, {\em Nesvorn\'y et al.} (2006) later revised the definition of the Karin cluster. In particular, they found that the original second-largest identified member of the family, (4507) 1990 FV, is in fact a background asteroid with no relation whatsoever to the recent breakup at the origin of the Karin cluster. Once this body is removed from the cluster membership, a large gap opens between the size of the largest family member and smaller members, a distribution that is now best reproduced in simulation by starting with a monolithic parent body. This change in implication for the internal structure of the parent body shows the importance of having a reliable estimate of the actual size distribution of family members. However, in the case of the Karin cluster, this change is problematic because the parent body of the Karin cluster is expected to be a rubble pile, if it is an original fragment of the Koronis-forming event. A solution proposed by {\em Nesvorn\'y et al.} (2006) is that the Karin cluster parent body was really formed by reaccumulation of smaller fragments during the Koronis family formation, as found in numerical simulations, but then, it was somehow consolidated into a more coherent body by various possible processes (lithification of regolith filling the interior, etc.). Another possibility is that in simulations, we are missing cases in which large intact fragments are created, so that the Karin cluster parent body could really have been a monolithic body. In fact, this systematic absence of large intact fragments in asteroid disruption simulations is often mentioned as a potential issue when discussing, for example, the internal structure of Eros, imaged by the NEAR/Shoemaker spacecraft (see Sec.\ 5.1).

A model of fragmentation adapted for microporous bodies (see the chapter by {\em Jutzi et al.} in this volume, for a definition of microporosity) has recently been implemented into an SPH hydrocode and tested against experiments on pumice targets ({\em Jutzi et al.}, 2008, 2009). It then became possible to simulate the formation of asteroid families from a microporous parent body. A microporous structure is assumed to be appropriate for parent bodies of dark taxonomic type or primitive asteroid families. In effect, several pieces of evidence point to the presence of a high degree of porosity in asteroids belonging to the C-complex, such as the low bulk density ($\approx 1.3$ g$/$cm$^3$)
estimated for some of them, for instance the asteroid (253) Mathilde encountered by the NEAR Shoemaker spacecraft ({\em Yeomans et al.}, 1997), and as inferred from meteorite analysis ({\em Britt et al.}, 2006). This model adapted for microporous bodies was used to reproduce the formation of the Veritas family, which is classified as a dark type family whose members have spectral characteristics of low-albedo, primitive bodies, from C to D taxonomic types ({\em Di Martino et al.}, 1997). This family is located in the outer main belt and is named after its apparent largest constituent, the asteroid (490) Veritas. The family age was estimated by two independent studies to be quite young, around 8 Myr ({\em Nesvorn\'y et al.}, 2003, {\em Tsiganis et al.}, 2007). Therefore, current properties of the family may retain signatures of the catastrophic disruption event that formed it. {\em Michel et al.} (2011) investigated the formation of the Veritas family by numerical simulations of catastrophic disruption of a 140-km-diameter parent body, which was considered to be the size of the original family parent body, made of either porous or non-porous material. Pumice material properties were used for the porous body, while basalt material properties were used for the non-porous body. Not one of these simulations was able to produce satisfactorily the estimated size distribution of real family members. Previous studies devoted to either the dynamics or the spectral properties of the Veritas family treated (490) Veritas as a special object that may be disconnected from the family. Simulations of the Veritas family formation were then performed representing the family with all members except Veritas itself. For that case, the parent body was smaller (112 km in diameter), and a remarkable match was found between the simulation outcome, using a porous parent body, and the real family. Both the size distribution and the velocity dispersion of the real reduced family were reproduced, while the disruption of a non-porous parent body did not reproduce the observed properties very well (see Fig.~2). This finding was consistent with the C spectral type of family members, which suggests that the parent body was porous and showed the importance of modeling the effect of porosity in the fragmentation process. It was then concluded that it is very likely that the asteroid (490) Veritas and probably several other small members do not belong to the family as originally defined, and that the definition of this family should be revised. This example shows how numerical modeling can better constrain the definition of (or the belonging to) an asteroid family, provided (i) that the fragmentation model used to simulate its formation is consistent with the possible material properties of the parent body, and (ii) that the family is young enough that a direct comparison with the modeling is possible.

\begin{figure*}
 \epsscale{1.}
 \plotone{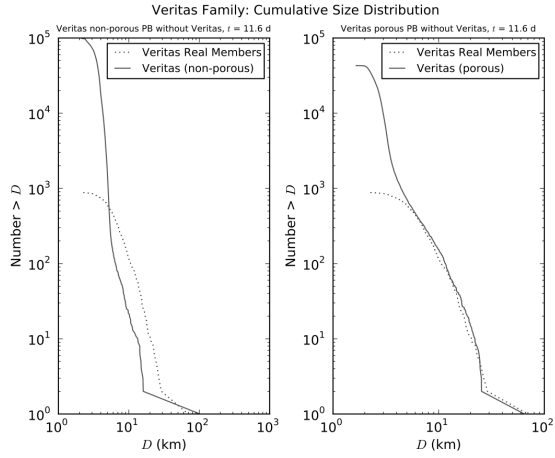}
 \caption{\small Cumulative size distributions of fragments from the simulations of the disruption of a Veritas monolithic parent body, either non-porous (left) or porous (right). The impact angle is $0^\circ$ (head-on) and the impact speed is 5 km/s. The distribution of the real family is also shown for comparison. In this case, the family consists of all members except Veritas itself, which reduces the size of the parent body to 112 km. The simulated time is about 11.6 days after the impact.}
 \end{figure*}

\bigskip
\noindent
\textbf{4.2 The Ejection Velocity Distribution of Family Members}
\bigskip

In addition to fragment sizes, numerical simulations also provide the ejection velocities. In general, impact simulations find that smaller fragments tend to have greater ejection speeds than larger ones. However, there is still a wide spread of values for fragments of a given mass, which makes it difficult to define a power-law relationship between fragment masses and speeds, such as the ones often used in collisional evolution models (see, e.g. {\em Davis et al.}, 2002). Figure 3 shows an example of this relation for a simulation reproducing the Eunomia family as a result of the disruption of a monolithic (basalt-like) parent body impacted at an impact angle of $45^\circ$.

\begin{figure*}
 \epsscale{1.}
 \plotone{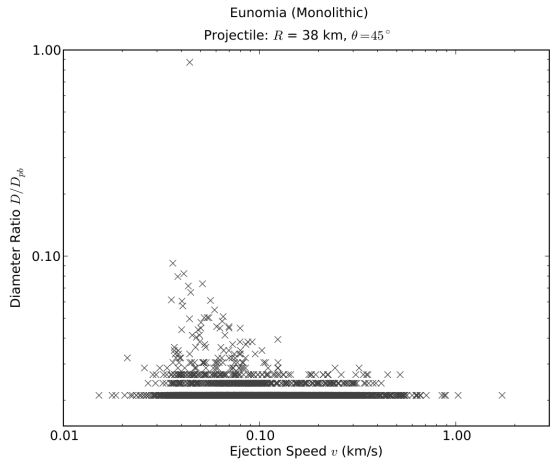}
 \caption{\small Fragment diameter $D$ (normalized to that of the parent body $D_{pb}$) vs.\ ejection speed in a log-log plot obtained from a monolithic Eunomia parent body simulation using a projectile impacting at an angle of incidence $\theta=45^\circ$. Only fragments with size above the resolution limit (i.e., those that underwent at least one reaccumulation event) are shown here. From {\em Michel et al.} (2004).}  
 \end{figure*}

In the case of real asteroid families, however, the dispersion of their members is characterized through their orbital proper elements, in particular their proper semimajor axis, eccentricity, and inclination. These elements have long been assumed to be essentially constants of motion that remain practically unchanged over astronomically long timescales (e.g., {\em Milani and Knezevi\'c}, 1994), although some perturbations have been found to be capable of modifying them, as we will explain below. Thus, we do not have direct access to the ejection velocities of family members. Fortunately, ejection velocities can be converted into a dispersion in orbital elements through Gauss' equations ({\em Zappal\`a et al.}, 1996), provided that both the true anomaly and the argument of perihelion of the family parent body at the impact instant are known or assumed.  For a given family, the estimated values of the barycenter semimajor  axis, eccentricity, and inclination can be used with the Gauss formulae up to first order in eccentricity to compute for each member the distance of its orbital elements $\delta a$, $\delta e$ and $\delta I$ from the barycenter of the family:  
\begin{equation} 
\scriptsize
\left\lbrace
\begin{array}{rcl}
\displaystyle \frac{\delta a}{a_b} &=& \displaystyle
\frac{2}{na_b\sqrt{1-e_b^2}}\lbrack (1+e_b \cos f_0)V_T + e_b \sin f_0 V_R
\rbrack ,\\   
\delta e &=& \displaystyle \frac{\sqrt{1-e_b^2}}{na_b}\left[ \frac{e_b + 2
  \cos f_0 +e_b \cos^2 f_0}{1+e_b \cos f_0}V_T + \sin f_0 V_R \right] ,\\
\delta I &=& \displaystyle \frac{\sqrt{1-e_b^2}}{na_b} \frac{\cos( \omega
  +f_0)}{1+e_b \cos f_0} V_W ,\\
\end{array}
\right.
\normalsize
\end{equation}
where $V_T$, $V_R$ and $V_W$ are the components of the ejection velocity in the along-track, radial, and out-of-plane directions, respectively, $n$ is the mean motion, $f_0$ is the true anomaly of the parent body at the instant of the break-up and $\omega$ is its argument of perihelion. Since these last two angles are not known, their values must be assumed. {\em Zappal\`a et al.} (1996) showed that the most sensitive angle is $f_0$. Assuming different values of this  angle changes the shape of the cluster containing the family members in orbital element space. In other words, it defines whether the break-up generates a family that is spread in semimajor axis, in eccentricity, or in  inclination. 

Thanks to this conversion, it is thus possible to assess the realism of a numerical simulation of a family formation by comparing the dispersion of family members and simulated fragments in the same space. Unfortunately, other mechanisms exist that, depending on the age of the considered family, can obscure the original dispersion of family members. In fact, once a family is created, its members are subjected to various perturbations. In particular, high-order secular resonances, mean-motion resonances even involving multiple planets ({\em Morbidelli and Nesvorn\'y}, 1999), and the Yarkovsky thermal effect ({\em Farinella and Vokrouhlick\'y}, 1999) have been shown to be capable of altering the proper elements. Therefore, while proper elements have been assumed conventionally to retain the memory of the disruption outcome conditions, these later studies demonstrated that this is not necessarily true, even for the proper semimajor axis in the case when the asteroid is small enough that Yarkovsky drift is effective (see {\em Bottke et al.}, 2002, {\em Vokrouhlick\'y et al.}, this volume). Depending on how old the family is, the current proper elements of family members cannot be interpreted as reflecting their starting conditions; rather, they must be seen as a result of such secular processes acting over time, whose effects are to cause a slow diffusion of family members in orbital-element space, starting from a smaller dispersion. The Koronis family is a good example showing these effects. The current distribution of Koronis family members in proper-element space is quite spread and its shape suggests that it has been subjected to the Yarkovsky effect as well as to the effects of nearby secular resonances and mean-motion resonances. {\em Bottke et al.} (2001) computed the dynamical evolutions of 210 simulated Koronis family members under the influence of the Yarkovsky effect and dynamical diffusion due to several resonances (namely, the 5:2 and 7:3 mean-motion resonances with Jupiter, a secular resonance that involves the precession rate of the small body's longitude of perihelion $g$ and the fundamental frequencies of Jupiter $g_5$ and Saturn $g_6$). The test family members were started with a dispersion that is consistent with the ones obtained from impact simulations of Koronis family formation. They were integrated over 700 Myr, which is still shorter than the estimated age of the family ($>1$ Gyr). However, this evolution showed that the current shape of the family cluster in proper-element space does not represent the original one from the collisional event but is well explained by its subsequent evolution.

Fortunately, if a family is young enough, its dispersion can still be close to the original one resulting from the parent body break-up, and in that case, the comparison between numerical simulations of family formation and actual family dispersion is straightforward. On the other hand, the degree of spreading observed now, together with the knowledge of the degree of dispersion resulting directly from the break-up by numerical simulations, can better constrain the age of the family, once the efficiency of the diffusion processes is well assessed.  {\em Nesvorn\'y et al.} (2002, 2003) identified several asteroid families with formation ages smaller than 10 Myr. These families represent nearly the direct outcome of disruptive asteroid collisions, because the observed remnants of such recent break-ups have apparently suffered limited dynamical and collisional erosion ({\em Bottke et al.}, 2005). As already described in the previous section, the Karin cluster and the Veritas family belong to this group of such young families. 

Fig. 4 shows an example in which the dispersion of the actual Veritas family is compared with that of fragments from an impact simulation of Veritas family formation. The simulated dispersion matches the shape of the ellipses representing the real dispersion. This result is consistent with the expectation that the orbital extent of the family is not produced by post-diffusion processes, which gives some credibility to family formation simulations that reproduce both the size distribution and velocity dispersion of actual members.

\begin{figure*}
 \epsscale{1.}
 \plotone{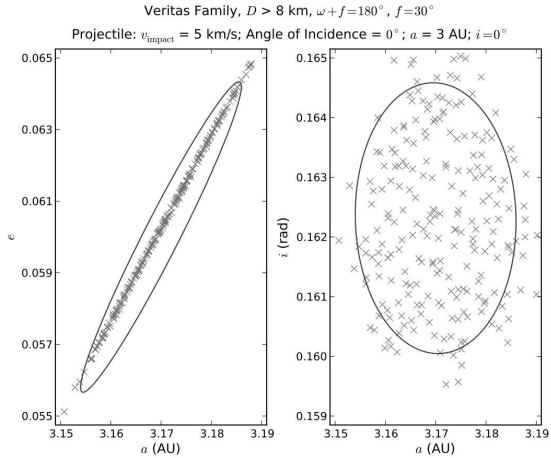}
 \caption{\small Distribution of fragments larger than 8 km from a simulation of disruption of a porous monolithic Veritas family parent body excluding Veritas itself as a result of a head-on impact of a projectile at 5 km/s with orbital semi-major axis $a$ of 3 AU and inclination $I$ of $0^\circ$. The outcome is represented in the aÐe plane (left) and $a$-$I$ plane (right). The superimposed ellipse is an equivelocity curve for speed cutoff of $40$ m$/$s, parent body true anomaly $f=30^\circ$ and argument of perihelion $\omega= 150^\circ$. This curve was defined by {\em Tsiganis et al.} (2007) as that closest to representing the dispersion of the real Veritas family in orbital element space. From {\em Michel et al.} (2011).}  
 \end{figure*}

\bigskip
\centerline{\textbf{ 5. IMPLICATIONS}}
\bigskip

\bigskip
\noindent
\textbf{5.1 Internal Structure of Asteroids}
\bigskip

According to numerical simulations of family formation, all fragments produced by the catastrophic disruption of a large asteroid (typically larger than 1 km in diameter, e.g. in the gravity regime) consist of self-gravitating aggregates, except the smallest ones. If this is correct, then most asteroids of at least second generation should be rubble piles. We note that {\em Campo Bagatin et al.} (2001) ran a number of simulations of main belt collisional evolution to assess the size range where reaccumulated bodies should be expected to be abundant in the main asteroid belt. They found that this diameter range goes from about 10 to 100 km, but may extend to smaller or larger bodies, depending on the prevailing collisional response parameters, such as the strength of the material, the strength scaling law, the fraction of kinetic energy of the impact transferred to the fragments, and the reaccumulation model.

The collisional lifetime of bodies larger than a few tens to hundreds of kilometers in diameter is longer than the age of the Solar System, suggesting that most bodies in that size range are likely to be primordial, while smaller bodies are probably collisional fragments (see, e.g., {\em Bottke et al.} 2005 and the chapter by Bottke et al.\ in this volume). The exact size above which a body is more likely to be primordial is somewhat model-dependent. {\em Binzel et al.} (1989), from a study of lightcurves, suggested that this transition occurs at a diameter of $\approx 125$ km. However, as this is a statistical measure, some smaller asteroids may still be primordial and some larger ones may have broken up in the past.  In fact, due to the variability in possible interior starting compositions, and variations in the chaotic dynamics of accumulation, the size above which a body is more likely to be primordial is dependent on the specific formation scenario, as well as the compositions, masses and velocities involved. Thus, some asteroids smaller than 100 km may still be primordial, and some larger ones may have broken up catastrophically in the past. This is especially true if one goes back to the very earliest formation, in the first few million years, when considering hit-and-run collisions (see the chapter by {\em Asphaug et al.} in this volume). These may have completely disrupted some of the largest asteroids, as projectiles, when they experienced grazing collisions into larger target embryos. This makes the internal structure of middle-sized asteroids one of the most important aspects of these bodies that can be determined by future space missions and observations, allowing us to test our interpretations based on theoretical collisional studies (see also {\em Scheeres et al.}, this volume).

During the past 4 billion years, catastrophic disruption has been the result of hypervelocity collisions. {\em Bottke et al.} (2005) estimate that about 20 asteroid families have formed from the breakup of parent bodies larger than 100 km diameter over the last 4 billion years. But several hundred asteroids currently exist in the 100 km size range, making it likely that most of these are original bodies. In this regard, asteroid 21 Lutetia, approximately 90 km diameter, is a scientifically important object, of which we have obtained a quick glimpse during Rosetta's 2010 flyby  (see Barucci et al., this volume). The relatively high measured mass (bulk density 3.1 g$/$cm$^3$) led {\em Weiss et al.} (2012) to interpret Lutetia as being a partly differentiated, impact-shattered, but largely intact parent body, covered in a predominately chondritic outer component. Other interpretations are of course possible.

Assuming that the transition between primordial and second-generation bodies occurs at diameters about 100 km, what about Eros, whose diameter is much below this threshold and therefore should be a fragment of a larger body? There is still a debate about the internal structure of this asteroid as the images of its surface can be explained by either a fractured (but solid/strength-dominated) structure or a rubble pile ({\em Asphaug} 2009, chapter by {\em Marchi et al.} in this volume). However, if Eros is not a rubble pile, its formation as a fragment of a large asteroid would need a solution that is not yet found in numerical simulations of catastrophic disruptions. Another point of view could thus be that it is a monolithic body that has been shattered in place (e.g. {\em Housen} 2009; {\em Buczkowski et al.} 2008). In this case major impacts fracture it in place, introducing only modest increases to its porosity. This requires very low strain rate of expansion, e.g. a small elastic strain at fracture, which may be consistent with size-dependent relationships for brittle failure. So this is probably feasible to form a {\it shattered monolith} when a single monolithic body is impacted, but with relatively low energy compared to disruption. But then one must ask where did the single monolithic body come from to begin with, and why has it not been subsequently fragmented and jumbled by slightly more energetic collisions? The alternative is that the grooves have nothing to do with brittle failure, but are instead planes of granular failure.

Thus, so far, the formation of a dispersed cloud of sizable fragments (larger than a few hundreds of meters) systematically requires that the parent body is first fragmented into small pieces, down to the resolution limit of simulations (a few hundreds of meters), and then that gravitational reaccumulation takes place to form larger final remnants. This is probably what happened for Itokawa, which appears to be a rubble pile ({\em Fujiwara et al.}, 2006). In fact, if the major blocks on Itokawa were intact monolithic, then they would give us a kind of lower size range of intact fragments produced from large impacts. Using the version of pkdgrav with enhanced collision handling to preserve shape and spin information of reaccumulated bodies ({\em Richardson et al.,} 2009),  {\em Michel and Richardson} (2013) showed that the process of catastrophic disruption and gravitational reaccumulation can form fragments with similar shapes as Itokawa's shape and can explain the presence of a large amount of boulders on the surface, as observed. Fig.~5 shows the outcome of such a simulation. We note that in this kind of modeling, the shapes of the aggregates formed by the reaccumulation process are parameter dependent. In particular, if we change the assumed strength of the aggregates or the bouncing coefficients (see {\em Michel and Richardson}, 2013, for a definition of these parameters), the final shape may be different. For instance, a preliminary simulation using a lower strength leads to a final largest aggregate that is more spherical. Because of the lower assumed strength, reaccumulating aggregates break more easily as a result of tidal and rotational forces, and therefore the object produced by this reaccumulation has difficulty keeping its irregular shape and instead becomes more and more rounded. Further studies are required to determine whether this type of outcome has some interesting implications, and to assess the actual sensitivity of the final shapes of reaccumulated objects to the parameters. It may be that we can provide some rough constraints on some of the mechanical properties of asteroids whose shapes are known, based on the parameters required to form them using this model. An extensive set of simulations is planned for this purpose, that will require long runs with current computer power.

 \begin{figure*}
 \epsscale{.3}
\plotone{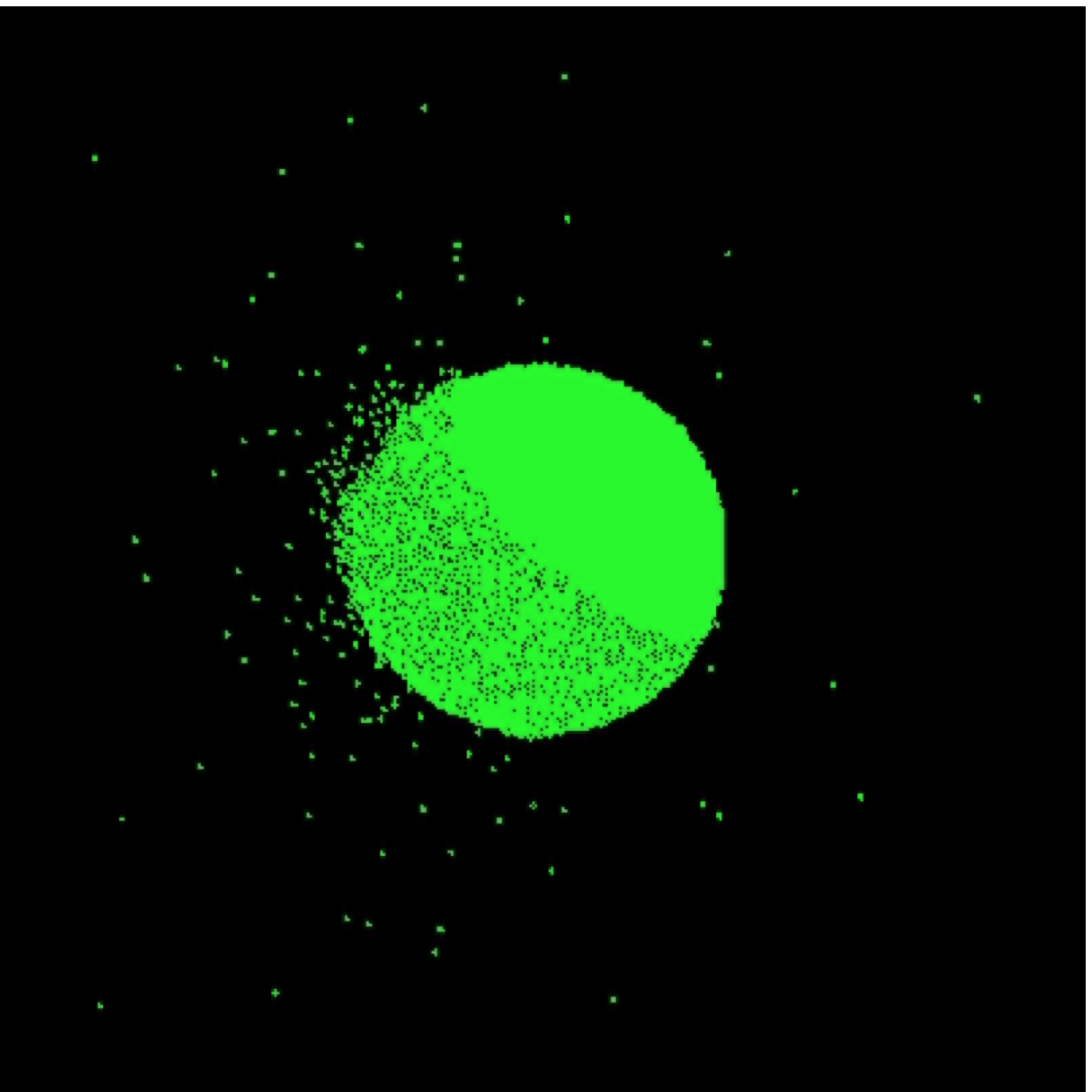}
\plotone{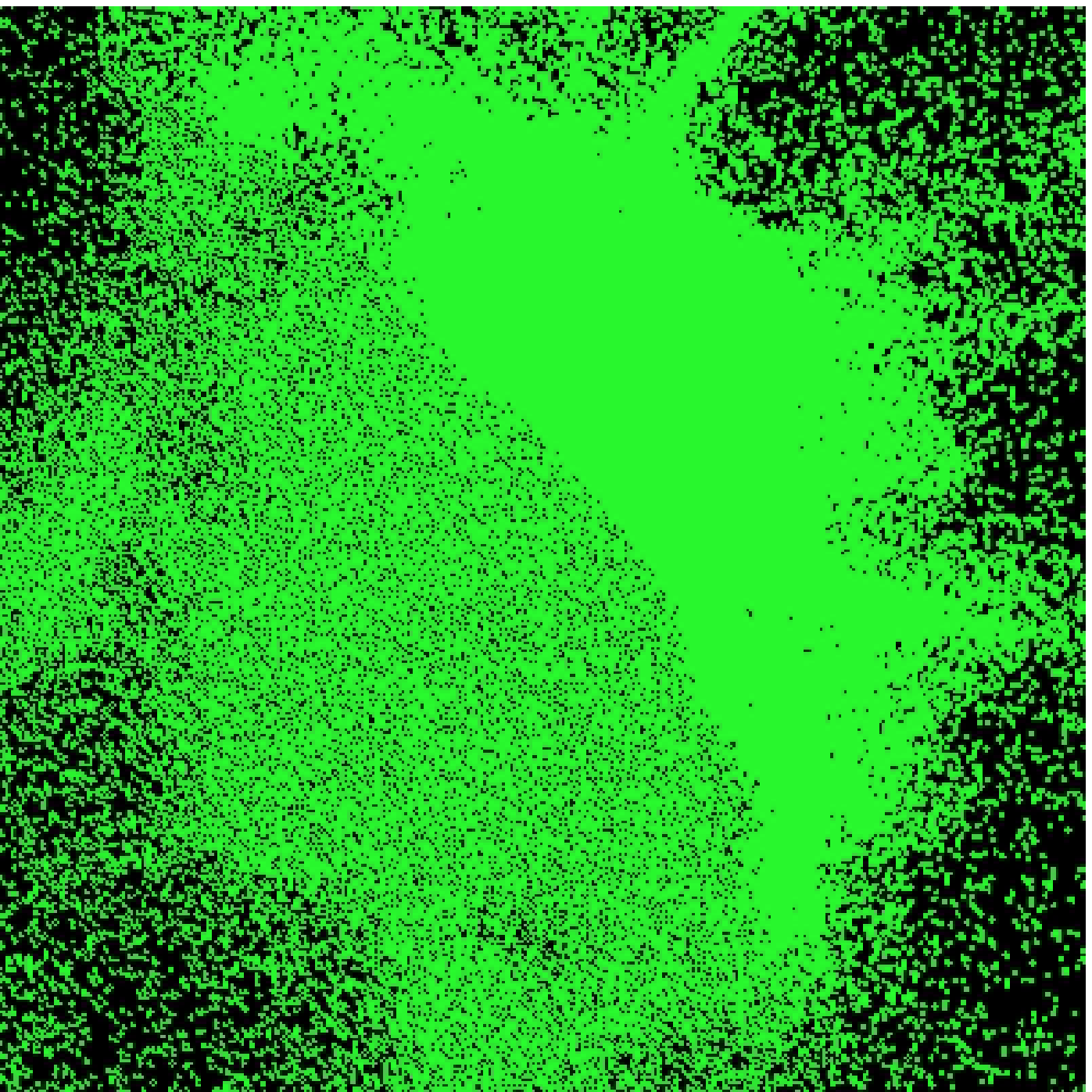}
\plotone{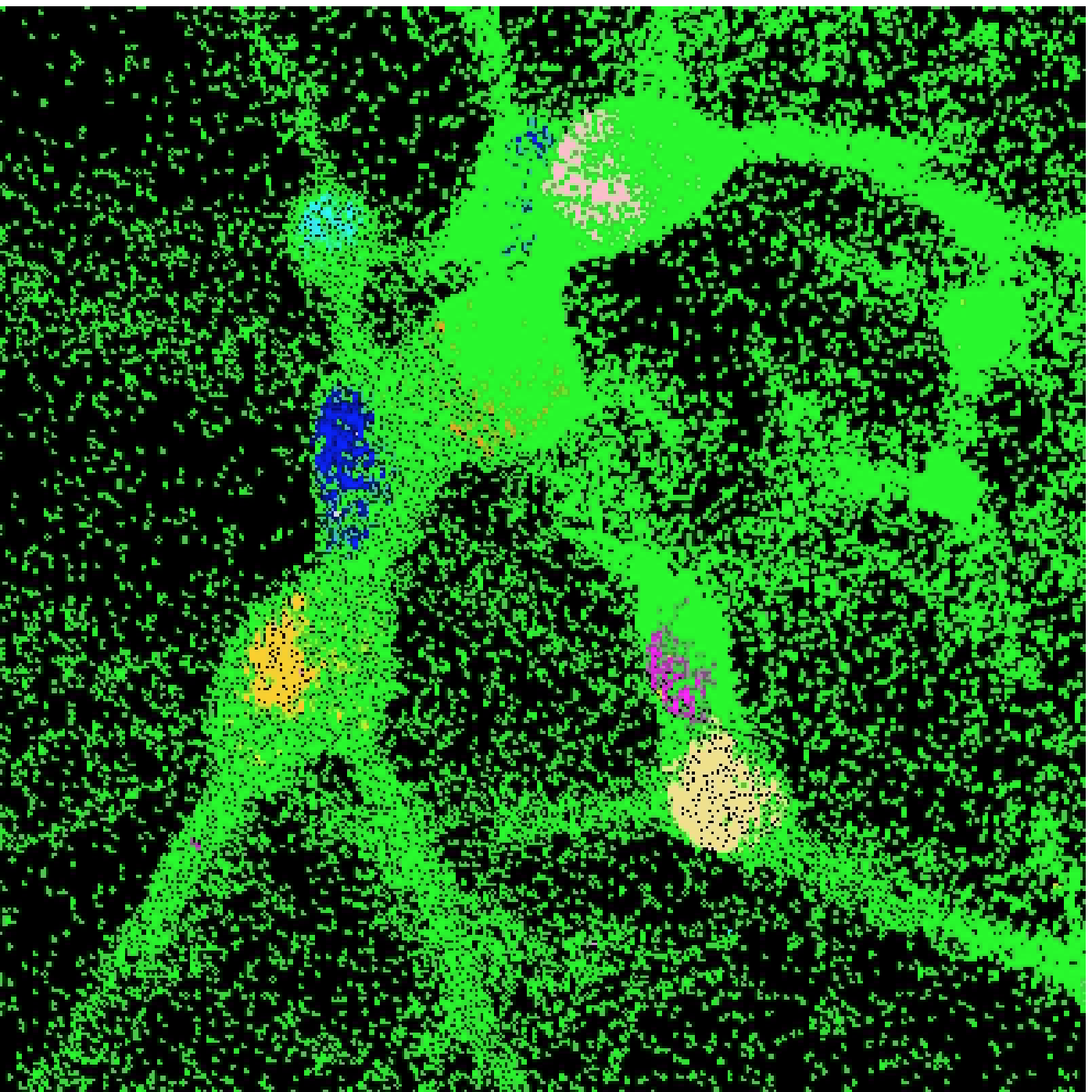}
\plotone{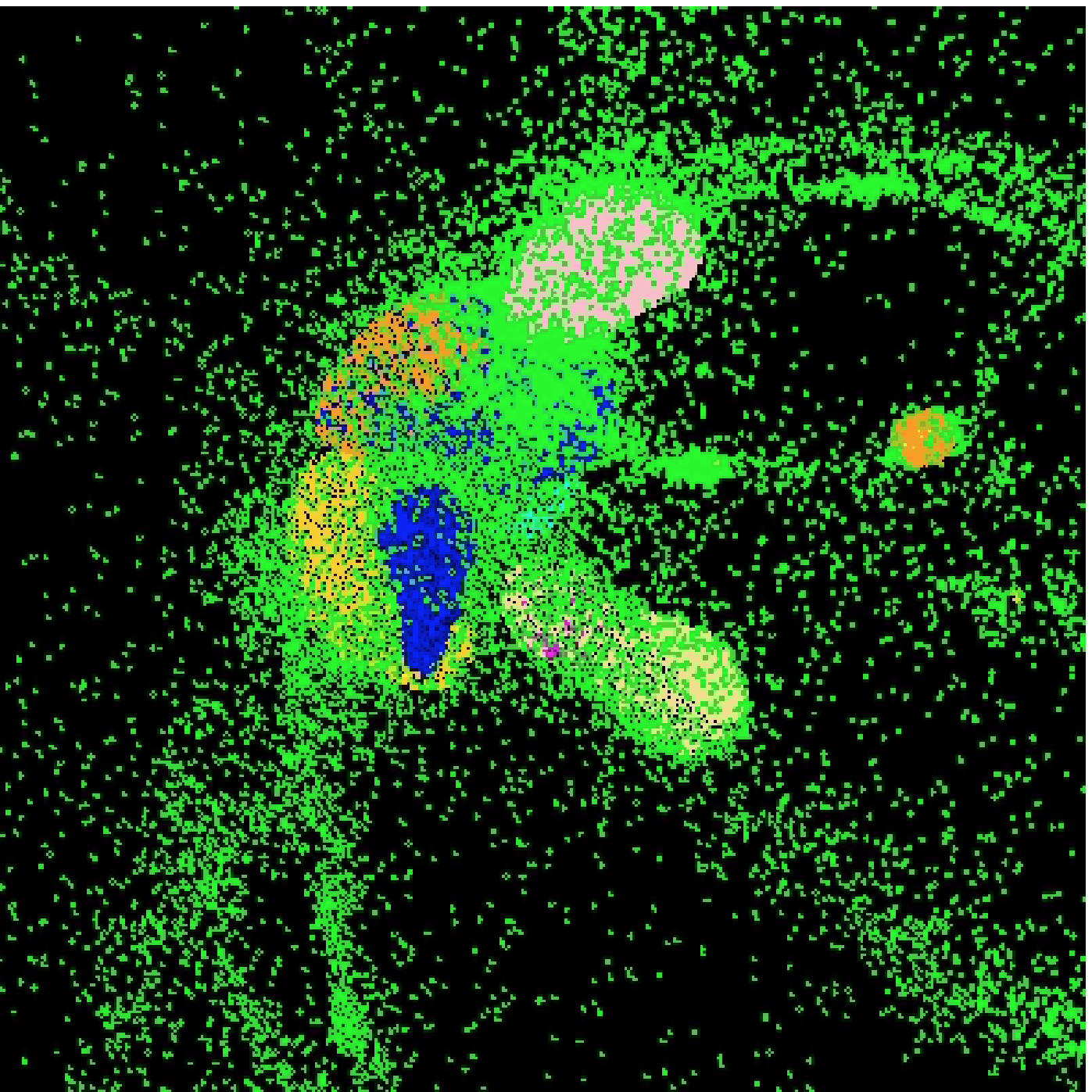}
\plotone{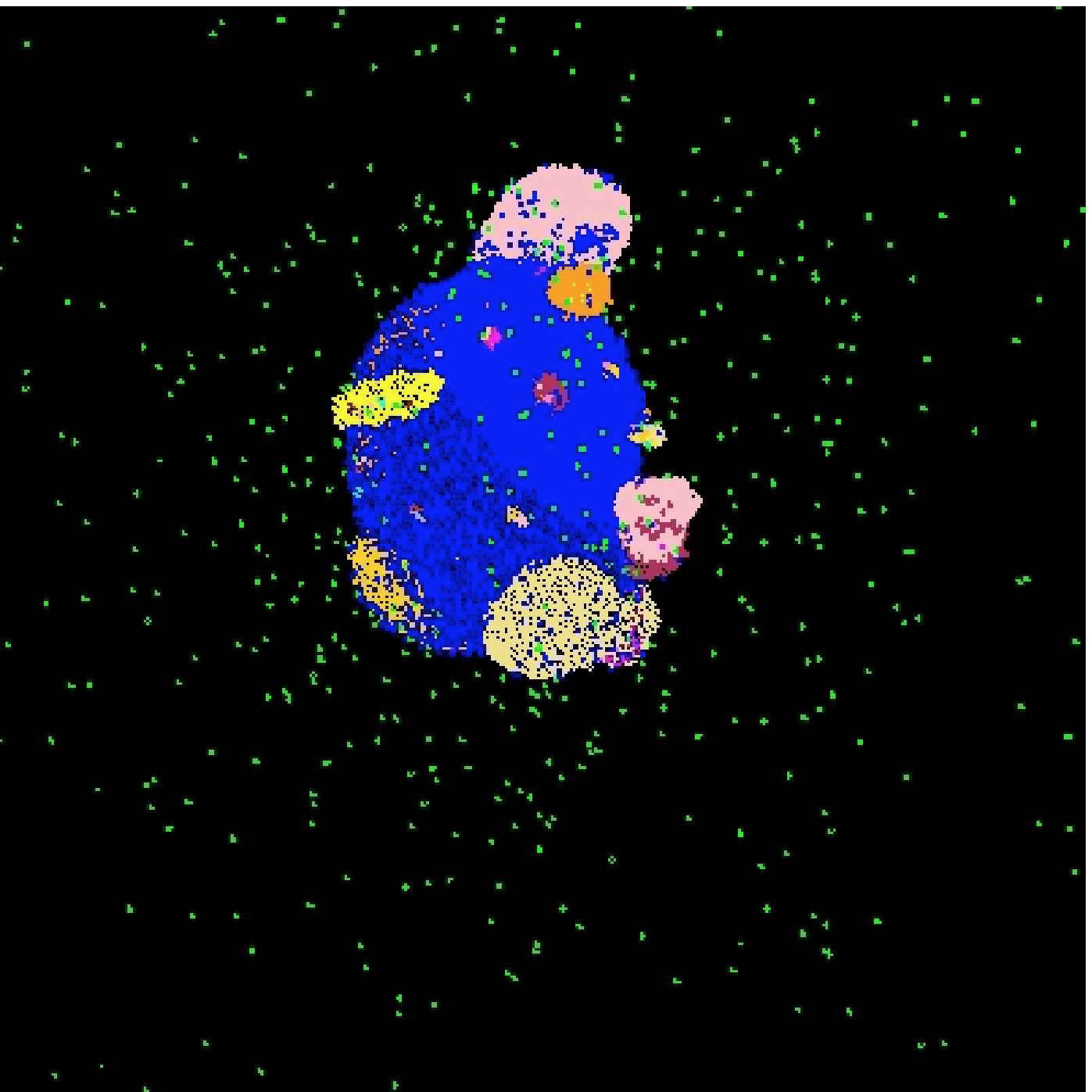}
\plotone{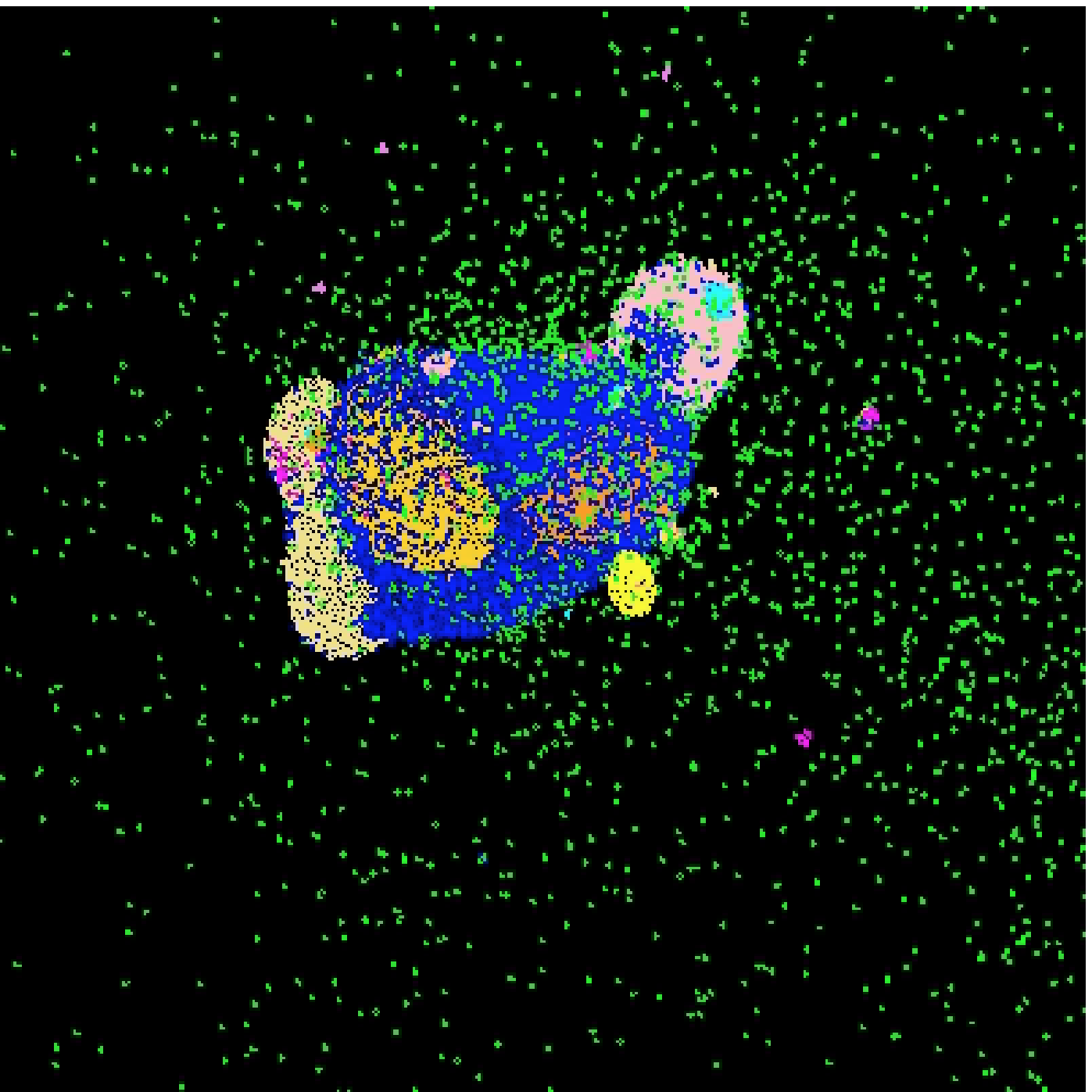}
 \caption{\small Snapshots of the reaccumulation process following the disruption of a 25 km diameter asteroid. From left to right: first instant at the end of the fragmentation phase when all fragments (white dots) are about 200 m in diameter; the ejection of those fragments a few seconds later; the first reaccumulations that occur because of the slow relative speed between some fragments, showing the formation of a few aggregates represented by different grey levels; the formation of the largest fragment of this disruption by reaccumulation of several aggregates into a single body; and the final largest fragment shown at two different instants: the boulders on its surface and its overall shape are reminiscent of Itokawa. Credit: Michel and Richardson, A \& A, 554, L1, 2013, reproduced with permission \copyright ESO.}  
 \end{figure*}

The production of rubble goes up with size, because it gets harder and harder (with increasing gravity) to liberate mass to escape speed than to beat it into small fragments that eventually can reaccumulate. The implication is that if Itokawa is a rubble pile, then Eros should be even more so. Whether this is represents reality awaits direct seismic or internal-structure exploration (e.g., by radar tomography) of objects Eros-sized and larger (see also {\em Scheeres et al.}, this volume).

\bigskip
\noindent
\textbf{5.2 Compositions}
\bigskip

Originally families were only identified on the basis of dynamical considerations. Then, once spectral observations became available, it was found that the vast majority of those families identified by dynamics showed remarkable homogeneous spectral properties (see {\em Masiero et al.}, this volume). So, homogeneity in terms of spectral properties seems to be the norm. However, when an object satisfies the distance criterion to be associated with a family, it is often considered as an interloper when its spectral properties do not match. Therefore, the identification of family membership also relies on homogeneous spectral properties and whether the homogeneity in spectral properties is a reality or an assumption is not clear yet. Such homogeneity can only be explained if the family parent body was homogenous itself, so that when fragments reaccumulate during the reaccumulation phase, there's no mixture of different materials taking place. Alternatively, it may also be that the reaccumulation process does not mix different materials that could be initially present in the parent body or mixes it so well that the outcome still looks homogeneous. Otherwise, if the parent body was heterogeneous in composition and if some mixtures happened, then the resulting family would show a variety of spectral properties within its members.  In fact, if re-accumulation is a random process, we expect the particles of a given large fragment to originate from uncorrelated regions within the parent body. In that case, if the parent body was heterogeneous in composition, then the composition of reaccumulated fragments could be a mixture of various material. Conversely, if the initial velocity field imposed by the fragmentation process determines the re-accumulation phase, the particles belonging to the same fragment should originate from well-defined areas inside the parent body. In addition, the position and extent of these regions would provide indications about the mixing occurring as a result of the re-accumulation process. 

{\em Michel et al.} (2004) traced back, at least for some of the largest fragments, the original positions within the parent body of the particles that end up forming the aggregates during some family formations. As an example, they traced the particles belonging to the three largest fragments of their simulation of the Koronis family formation back to their original positions inside the parent body. Recall that the Koronis family was formed in a highly catastrophic event, as its largest member is estimated to contain only $4 \%$ of the parent body's mass. In such an event, the re-accumulation process lasts up to several days, much longer than for a barely disruptive event, and gives rise to many gravitational encounters. Therefore, this kind of event may well lose the memory of the initial velocity field. Nevertheless, it was found that particles forming a large reaccumulated fragment originate from well-clustered regions within the parent body. This indicates that re-accumulation is definitely not a random process. Interestingly, the position of the original region depends greatly on the internal properties of the parent body. The largest remnant of the pre-shattered model of the Koronis parent body involves particles that were initially located between the core and the region antipodal to the impact point. Conversely, in the monolithic parent body, those particles were initially much more clustered in the core region, with no particles originating from the antipode. This difference also holds true for the next largest fragments. Nevertheless, these results indicate that the velocity field arising from the fragmentation phase has a major influence on the reaccumulation process. Particles that eventually belong to a given fragment originate from the same region inside the parent body. However, this location (as well as its extent, which determines the degree of mixing of the fragments) depends also on the parent body's internal properties in a complex way. Recently {\em Michel et al.} (2015) looked at the cases of parent bodies with internal structures that could represent large asteroids formed early in the Solar System history. Some results are shown in {\em Jutzi et al.} (this volume). They confirm that most particles in each reaccumulated fragments are sampled from the same original region within the parent body. However, they also found that the extent of the original region varies considerably depending on the internal structure of the parent body and seems to shrink with its solidity.

As a conclusion, the spectral homogeneity within a family may represent the material homogeneity of the initial parent body. It may also be due to the way reaccumulation takes place. But in that case and if the parent body was heterogeneous, although each family member would still be homogeneous, we may expect different spectral properties from one member to the other, depending on which original region of the parent body it samples. The fact that most families do not show strong spectral variations between family members, at least in the data of ground-based observations and except if this is imposed by the membership criterion, is consistent with the idea of homogeneity of the family parent body.

\bigskip
\centerline{\textbf{ 6. CONCLUSIONS}}
\bigskip

Our understanding of the collisional physics and our account for gravity in large asteroid disruptions have allowed numerical simulations to reproduce successfully the formation of asteroid families, in agreement with the idea that these families originate from the disruption of a large parent body. Simulation results indicate that asteroid family members are not just the product of the fragmentation of the parent body, leading to intact fragments, but rather the outcome of the subsequent gravitational phase of the event, which allows some of the intact fragments to reaccumulate and form gravitational aggregates, or rubble piles. For all considered cases so far (family parent bodies of diameter typically larger than tens of kilometers), this outcome is systematic for fragments larger than a few hundred meters. Thus, according to simulations, since it is believed that most bodies smaller than $100$ km in diameter originate from the disruption of a larger body, then they should be rubble piles or heavily shattered bodies, which is consistent with the low measured bulk densities for some of them and the finding by {\em Campo Bagatin et al.} (2001) based on main belt collisional evolution modeling. Therefore, exploring the origin of asteroid families unexpectedly led to a result that has great implications for the entire asteroid population and its history. 

Moreover, it was also found that the outcome of a disruption is very sensitive to the original internal structure of the parent body, in particular the kind and amount of internal porosity. Thus, the comparison between simulation outcomes for various kinds of parent-body structures (monolithic, pre-shattered, microporous, rubble pile) and real family properties can help to constrain the internal properties of the parent body of the considered families and in the family identification itself. For instance, it was found that the Veritas family is very well reproduced if the asteroid Veritas itself is excluded from its family, which was already recognized as a possibility before disruption simulations were performed to model the formation of this family. 

Thanks to the improved sensitivity of observations, allowing us to reach smaller asteroid sizes, and to the tools developed to better define asteroid families, new asteroid families keep being identified, especially small and young ones. The latter, which have not been affected yet by dynamical diffusion or post-collisional processes, are a good test for numerical simulations that must be able to reproduce them as they are. Such an exercise, which has already been done successfully for some young families (e.g., Veritas, Karin) must keep being performed as a check for our numerical models. In particular, new fragmentation models are continuously being developed, accounting for various possible strength models and fragmentation modes. Once they are validated at small laboratory scales by comparison with impact experiments, they can be used at large scale (with an associated $N$-body code) to reproduce young family properties, allowing us to increase the range of internal structures and fragmentation modes that can be considered for the parent body. This modeling work, calling for different models, is crucial to better constrain the possible internal structure of family parent bodies, to refine the definition of a family, and to understand whether some families are formed from differentiated/heterogeneous parent bodies, despite their apparent (or assumed) homogeneities. Asteroid families are very important tracers of the entire asteroid belt history and as we have already seen, their understanding can have profound implications on our determination of the physical properties of asteroids in general. 

More work is also required to check in which context large intact fragments can be produced in numerical simulations of large asteroid disruptions. Although there is no firm conclusion about Eros' internal structure, the fact that it may be a shattered object only and not a rubble pile raises the issue of the formation of such large fragments in a collision. It may also be that the reaccumulation process is followed by internal processes that may consolidate boulders together. Such processes would eventually transform reaccumulated fragments into a coherent body. If this were the case, then reaccumulation would not necessarily imply a rubble-pile structure. Simulations of the reaccumulation phase now include the possibility to account for the final possible shapes of reaccumulated fragments. This modeling needs further improvement to  increase its realism but it will be very difficult, if even possible, to achieve the level of complexity needed to model the internal processes that may consolidate boulders together. Asteroid internal processes are poorly understood and depend on too many parameters and unknowns. 

Space missions dedicated to direct measurement of internal structures, and possibly to their response to an impact (e.g., by using a kinetic impactor), are thus crucial to improve our understanding of these important internal properties of asteroids and to check our modeling of the collisional and internal processes. Moreover, sample return missions as well as visits/fly-bys of members of asteroid families would also provide detailed information on their physical properties and would allow us to check whether ground-based measurements wash out some important data regarding their composition and possible variations among members. Asteroid families and the collisional process, which is at the heart of family formation and evolutionary main belt history, rely on our efforts to combine complex models and space- ground-based measurements. 

\textbf{ Acknowledgments.} We are grateful to Clark Chapman and an anonymous reviewer for their comments that greatly helped to improve the chapter. P. M. acknowledges support from the French space agency CNES and the French national program of planetology. D.C.R. acknowledges NASA grant NNX08AM39G and NSF grant AST1009579 (and previous NASA/NSF grants). D. D. D. acknowledges support from the National Science Foundation (Planetary Astronomy Program grants AST0098484, AST0407045, and AST0708517). M. J. is supported by the Ambizione program of the Swiss National Science Foundation.

\bigskip

\centerline\textbf{ REFERENCES}
\bigskip
\parskip=0pt
{\small
\baselineskip=11pt

\refs Arnold, J. R. (1969) Asteroid families and jet streams. {\em Astron. J., 74}, 1235-1242.

\refs Asphaug, E., Ryan, E. V., and Zuber, M. T. (2002) Asteroid interiors. In {\em Asteroids III} (W. F. Bottke Jr., A. Cellino, P. Paolicchi., and R. P., Binzel, eds.), pp. 463-484, Univ. of Arizona Press, Tucson.

\refs Asphaug, E. (2009) Growth and Evolution of Asteroids. {\em Annual Review of Earth and Planetary Sciences, 37}, 413-448.

\refs Ballouz, R. -L., Richardson, D. C., Michel, P., and Schwartz, S. R. (2014a) Rotation-dependent catastrophic disruption of gravitational aggregates. {\em Astrophys. J., 789}, 158.

\refs Ballouz, R. -L., Richardson, D. C., Michel, P., Schwartz, S. R., and Yu, Y. (2014b) Numerical Simulations of Collisional Disruption of Rotating Gravitational Aggregates: Dependence on Material Properties. \planss, accepted.

\refs Benavidez, P., Durda, D. D., Enke, et al. (2011) A comparison between rubble-pile and monolithic targets in impact simulations: Application to asteroid satellites and family size distributions. {\em Icarus, 219}, 57-76.

\refs Benz, W., and Asphaug, E. (1994) Impact simulations with fracture. I. Method and tests. {\em Icarus, 107}, 98-116.

\refs Benz, W., and Asphaug, E. (1995) Simulations of brittle solids using smooth particle hydrodynamics. {\em Comput. Phys. Comm., 87}, 253-265.

\refs Benz,W., and Asphaug, E. (1999) Catastrophic disruptions revisited. {\em Icarus, 142}, 5-20.

\refs Binzel R. P., Farinella P., Zappal\`a, V, and Cellino, A (1989) Asteroid rotation rates-distributions and statistics. In {\em Asteroids II} (R. P., Binzel, T., Gehrels, M. S., Matthews, eds.), University of Arizona Press, Tucson, 416-441.

\refs Bottke, W. F., Vokrouhlick\'y, D., Borz, M., Nesvorn\'y, D., and Morbidelli, A. (2001) Dynamical spreading of asteroid families via the Yarkovsky effect. {\em Science, 294}, 1693Ð1696.

\refs Bottke, W. F., Vokrouhlick\'y, D., Rubincam, D. P., and Broz, M. (2002) The effect of Yarkovsky thermal forces on the dynamical evolution of asteroids and meteoroids. In {\em Asteroids III}  (W. F. Bottke, A., Cellino, P., Paolicchi, and R. P., Binzel, eds.), Univ. of Arizona Press, Tucson, 395-408.

\refs Bottke, W. F., Durda, D. D., Nesvorn\'y, D., Jedicke, R., Morbidelli, A., Vokrouhlick\'y, D., and Levison H. (2005) The fossilized size distribution of the main asteroid belt. {\em Icarus, 175}, 111-140.

\refs Britt, D.T., Consolmagno, G.J., and Merline, W.J. (2006) Small body density and porosity: New data, new insights. {\em Lunar Planet. Sci., 37}, abstract 2214.

\refs Buczkowski, D.~L., Barnouin-Jha, O.~S., Prockter, L.~M. (2008) 433 Eros lineaments: Global mapping and analysis. {\it Icarus, 193}, 39-52.

\refs Campo Bagatin, A., and Petit, J.- M. (2001) Effects of the geometric constraints on the size distributions of debris in asteroidal fragmentation. {\em Icarus, 149}, 210-221.

\refs Campo Bagatin, A., Petit, J-M., and Farinella, P. (2001) How Many Rubble Piles Are in the Asteroid Belt? {\em Icarus, 149}, 198-209.

\refs Chapman, C. R., Davis, D. R., and Greenberg, R. (1982) Apollo asteroids: Relationships to main belt asteroids and meteorites. {\em Meteoritics, 17}, 193Ð194.

\refs Chapman, C. R., Paolicchi, P., Zappal\`a, V., Binzel, R. P., and Bell J. F. (1989) Asteroid families: Physical properties and evolution. In {\em Asteroids II} (R. P. Binzel, T. Gehrels, and M. S. Matthews, eds.), pp. 386-415, Univ. of Arizona, Tucson.

\refs Davis, D. R., Chapman, C. R., Greenberg, R., Weidenschilling, S. J., and Harris, A. W. (1979) Collisional evolution of asteroids - Populations, rotations, and velocities. In {\em Asteroids} (T. Gehrels ed.), pp. 528-557, Univ. of Arizona Press, Tucson.

\refs Davis, D. R., Chapman, C. R., Greenberg, R. , and Weidenschilling, S. J. (1985) . Hirayama Families: Chips Off the Old Block or Collections of Rubble Piles? {\em Bulletin of the American Astronomical Society, 14}, 720.

\refs Davis, D. R., Durda, D. D., Marzari, F., Campo Bagatin, A., and Gil-Hutton, R. (2002) Collisional evolutions of small body populations. In {\em Asteroids III} (W. F. Bottke Jr., A. Cellino, P. Paolicchi., and R. P., Binzel, eds.), pp. 545-558, Univ. of Arizona Press, Tucson.

\refs Di Martino, M., Migliorini, F., Zappal\`a, V., Manara, A., and Barbieri, C. (1997) Veritas asteroid family: Remarkable spectral differences inside a primitive parent body. {\em Icarus, 127}, 112-120.

\refs Durda, D. D., Bottke, W. F., Enke, B. L., Merline, W. J., Asphaug, E., Richardson, D. C., and Leinhardt, Z. M. (2004) The formation of asteroid satellites in large impacts: Results from numerical simulations. {\em Icarus, 170}, 243Ð257.

\refs Durda, D. D., Bottke, W. F., Nesvorn\'y, D., Enke, B. L., Merline, W. J., Asphaug, E., and Richardson, D. C. (2007) SizeÐfrequency distributions of fragments from SPH/N- body simulations of asteroid impacts: Comparison with observed asteroid families. {\em Icarus, 186}, 498Ð516.

\refs Durda, D. D., Movshovitz, N., Richardson, D. C., Asphaug, E., Morgan, A., Rawlings, A. R., and Vest, C. (2011) Experimental determination of the coefficient of restitution for meter-scale granite spheres. {\em Icarus, 211}, 849-855.

\refs Durda, D. D., Richardson, D. C., Asphaug, E., and Movshovitz, N. (2013) Size dependence of the coefficient of restitution: Small-scale experiments and the effects of rotation. {\em Lunar Plan. Sci., XLIV}, abstract no. 2263.

\refs Farinella, P., Davis, D. R., and Marzari, F. (1996) Asteroid families, old and young. In {\em Completing the Inventory of the Solar System} (T.W. Rettig and J. M. Hahn, eds.), {\em ASP Conference Series, 107}, pp. 45Ð55, ASP, San Francisco.

\refs Farinella, P., and Vokrouhlick\'y, D. (1999) Semimajor axis mobility of asteroidal fragments. {\em Science, 283}, 1507-1510.

\refs Fujiwara, A., Kawaguchi, J., Yeomans, D. K., et al. (2006) The rubble-pile asteroid Itokawa as observed by Hayabusa. {\em Science, 312}, 1330-1334.

\refs Hirayama, K. (1918) Groups of asteroids probably of common origin. {\em Astron. J., 31}, 185-188.

\refs Housen, K. (2009) Cumulative damage in strength-dominated collisions of rocky asteroids: Rubble piles and brick piles. \planss{\em, 57}, 142-153.

\refs Jaeger, J. C., and Cook, N. G. W. (1969) {\em Fundamentals of Rock Mechanics}. Chapman and Hall, London.

\refs Jutzi, M., Benz, W., and Michel, P. (2008) Numerical simulations of impacts involving porous bodies. I. Implementing sub-resolution porosity in a 3D SPH hydrocode. {\em Icarus, 198}, 242-255.

\refs Jutzi, M., Michel, P., Hiraoka, K., Nakamura, A. M., and Benz, W. (2009) Numerical simulations of impacts involving porous bodies. II. Confrontation with laboratory experiments. {\em Icarus, 201}, 802-813.

\refs Jutzi, M., Michel, P., Benz, W., and Richardson, D. C. (2010) The formation of the Baptistina family by catastrophic disruption: porous versus non-porous parent body. {\em Meteoritics Planet. Sci., 44}, 1877-1887.

\refs Jutzi, M. (2015) SPH calculations of asteroid disruptions: the role of pressure dependent failure models. \planss, in press. 

\refs Leinhardt, Z. M., and Stewart, S. T. (2009) Full numerical simulations of catastrophic small body collisions. {\em Icarus, 199}, 542-559.

\refs McGlaun, J. M., Thompson, S. L., and Elrick, M. G. (1990) CTH: A 3-dimensional shock-wave physics code. {\em Int. J. Impact Eng., 10}, 351-360.

\refs Marzari, F., Davis, D., and Vanzani, V. (1995) Collisional evolution of asteroid families. {\em Icarus, 113}, 168-187.

\refs Michel, P., Benz, W., Tanga, P., and Richardson, D. C. (2001) Collisions and gravitational reaccumulation: forming asteroid families and satellites. {\em Science, 294}, 1696-1700.

\refs Michel, P., Benz, W., Tanga, P., and Richardson, D. C. (2002) Formation of Asteroid Families by Catastrophic Disruption: Simulations with Fragmentation and Gravitational Reaccumulation. {\em Icarus, 160}, 10-23.

\refs Michel, P., Benz, W., and Richardson, D. C. (2003) Fragmented parent bodies as the origin of asteroid families. \nat, {\em 421}, 608-611.

\refs Michel, P., Benz, W., and Richardson, D. C. (2004) Disruption of pre-shattered parent bodies. {\em Icarus, 168}, 420-432. 

\refs Michel, P., Jutzi, M., Richardson, D. C., and Benz, W. (2011) The asteroid Veritas: an intruder in a family named after it? {\em Icarus, 211}, 535-545.

\refs Michel, P., and Richardson, D.C. (2013) Collision and gravitational reaccumulation: Possible formation mechanism of the asteroid Itokawa. \aap, {\em 554}, L1-L4.

\refs Michel, P., Jutzi, M., Richardson, D. C., Goodrich, C. A., Hartmann, W. K., and O'Brien, D. P. (2015) Selective sampling during catastrophic disruption:
Mapping the location of reaccumulated fragments in the original parent body. \planss, in press.

\refs Milani, A., and Knezevi\'c, Z. (1994) Asteroid proper elements and the dynamical structure of the asteroid belt. {\em Icarus, 107}, 219-254.

\refs Morbidelli, A., and Nesvorn\'y, D. (1999) Numerous weak resonances drive asteroids toward terrestrial planets orbits. {\em Icarus, 139}, 295-308.

\refs Nesvorn\'y, D., Bottke, W. F., Dones, L., and Levison, H. F.  (2002) The recent breakup of an asteroid in the main-belt region. {\em Nature, 417}, 720-771.

\refs Nesvorn\'y, D., Bottke, W. F., Levison, H. F., and Dones, L. (2003) Recent origin of the Solar System dust bands. {\em Astrophys. J.}, {\em 591}, 486-497.

\refs Nesvorn\'y, D., Enke, B. L., Bottke, W. F., Durda, D. D., Asphaug, E., and Richardson, D. C. (2006) Karin cluster formation by asteroid impact. {\em Icarus, 183}, 296Ð311.

\refs Richardson, D. C. (1994) Tree code simulations of planetary rings. {\em Mon. Nat. R. Astron. Soc., 269}, 493Ð511.

\refs  Richardson, D. C., Quinn, T., Stadel, J., and Lake, G. (2000) Direct large-scale N-body simulations of planetesimal dynamics. {\em Icarus, 143}, 45-59.

\refs Richardson, D. C., Leinhardt, Z. M., Melosh, H. J., Bottke Jr., W. F., and Asphaug, E. (2002) Gravitational aggregates: Evidence and evolution. In {\em Asteroids III} (W. F. Bottke Jr., A. Cellino, P. Paolicchi., and R. P., Binzel, eds.), pp. 501-515, Univ. of Arizona Press, Tucson.

\refs Richardson, D. C., Michel, P., Walsh, K. J., and Flynn, K. W. (2009) Numerical simulations of asteroids modelled as gravitational aggregates with cohesion. \planss, {\em 57}, 183-192.

\refs Ryan, E. V., and Melosh, H. J. (1998) Impact fragmentation: From the laboratory to asteroids. {\em Icarus, 133}, 1-24.

\refs Schwartz, S.R., Richardson, D.C., and Michel, P. (2012) An Implementation of the Soft-Sphere Discrete Element Method in a High-Performance Parallel Gravity Tree-Code. {\it Granular Matter}, DOI 10.1007/s10035-012-0346-z.

\refs Tanga, P., Cellino, A., Michel, P., Zappal\`a, V., Paolicchi, P., and Dell'Oro, A. (1999) On the size distribution of asteroid families: the role of geometry. {\em Icarus, 141}, 65-78.

\refs Tillotson, J. H. (1962) Metallic Equations of State for Hypervelocity Impact. {\em General Atomic Report GA-3216,} July 1962.

\refs Tsiganis, K., Knezevi\'c, Z., and Varvoglis, H. (2007) Reconstructing the orbital history of the Veritas family. {\em Icarus, 186}, 484-497.

\refs Weibull, W. A. (1939) A statistical theory of the strength of material (transl.). {\em Ingvetensk. Akad. Handl., 151}, 5-45.

\refs Weiss, B. P., Elkins-Tanton, L. T., Barucci, M. A., et al. (2012) Possible evidence for partial differentiation of asteroid Lutetia from Rosetta. \planss, {\em 66}, 137-146.

\refs Yeomans, D. K., and 12 colleagues (1997) Estimating the mass of Asteroid 253 Mathilde from tracking data during the NEAR flyby. {\em Science, 278}, 2106-2109.

\refs Zappal\`a, V., Cellino, A., DellÕOro, A., Migliorini, F., and Paolicchi, P. (1996) Re-constructing the original ejection velocity fields of asteroid families. {\em Icarus, 124}, 156-180.

\end{document}